\documentclass[epj]{svjour}
\usepackage{graphics}
\newcommand{\spur}[1]{\not\! #1 \,}
\newcommand{\be}{\begin{equation}}
\newcommand{\ee}{\end{equation}}
\newcommand{\bea}{\begin{eqnarray}}
\newcommand{\eea}{\end{eqnarray}}
\begin{document}
\title{Nonfactorizable effects in B to charmonium decays}
\subtitle{}
\author{Fulvia De Fazio
}                     
\institute{Istituto Nazionale di Fisica Nucleare - Sezione di
Bari, Italy}
\date{Received: date / Revised version: date}
%
\abstract{ Nonleptonic $B$ to charmonium decays generally deviate
from the factorization predictions.  We study rescattering effects
mediated by intermediate charmed mesons
 in this class of decay modes and, in particular, we consider
 $B^- \to K^- h_c$ with
$h_c$ the $J^{PC}=1^{+-}$ $\bar c c$ meson, relating this mode to
$B^- \to K^- \chi_{c0}$. We find ${\cal B}(B^- \to K^- h_c)$ large
enough to be measured at the $B$ factories, hence this process
could be used to study the poorly known $h_c$.
\PACS{$12.39.$Hg\and$12.39.$St\and$13.25.$Hw} 
} 
\maketitle
\section{Introduction}
\label{intro} Testing the Standard Model description of CP
violation in the $B$ sector requires  a reduced theoretical
uncertainty,  a difficult task for nonleptonic  decays since a
general computational scheme has still to be developed. The
amplitude of two-body nonleptonic $B$ decays is given by the
matrix element of the effective Hamiltonian for $B \to M_1 M_2$
\cite{Buras:1998rb}: \be A(B \to M_1 M_2) = {G_F \over \sqrt 2}
\sum_i \lambda_i c_i(\mu) \langle M_1 M_2 | {\cal O}_i(\mu) | B
\rangle \,. \label{twobody} \ee  $\lambda_i$ are CKM matrix
elements, $c_i(\mu)$ Wilson coefficients evaluated at the scale
$\mu$ and ${\cal O}_i$ four-quark operators. The naive
factorization ansatz expresses their matrix elements as products
of matrix elements of quark currents.
\par
Let us consider $B^- \to K^- M_{\bar c c}$, where  $M_{\bar c c}$
belongs to the charmonium system. Neglecting the annihilation term
(suppressed by $V_{ub}$), the factorized amplitude is: \bea {\cal
A}_F(B^- \to K^- M_{\bar c c}) &=& {G_F \over \sqrt 2} V_{cb}
V^*_{cs} \Big[a_2(\mu)+\sum_{i=3,5,7,9}
a_i(\mu) \Big] \nonumber \\
\langle K^-|(\bar s b)_{V-A}|B^-\rangle &\times& \langle M_{\bar c
c} |(\bar c c)_{V \mp A}|0\rangle \label{matrixel} \eea with
$a_2=c_2+{c_1/N_c}$, $a_i=c_i+{c_{i+1}/N_c}$. Eq.(\ref{matrixel})
shows the drawbacks of the approach: scale and scheme dependence
of the  $c_i(\mu)$ are no more compensated by  that of the matrix
elements; besides, the product of these does not contain any
strong phase.  An improvement consists in the generalized
factorization, with $a_i(\mu)$ replaced by effective (process
independent) parameters $a_i^{\rm eff}$  to be fixed using
experimental data.  Other methods, like QCD-improved factorization
\cite{Beneke:2000ry}, PQCD \cite{Keum:2000ph}, SCET
\cite{Bauer:2001cu}, QCD sum rules \cite{Colangelo:2000dp}, can
only be applied to selected classes of nonleptonic modes.
\par
Generalized factorization indicates nonfactorizable effects in $B
\to K^- J/\psi$. The experimental ${\cal B}(B \to K^- J/\psi)$ can
be fitted with  $|a_2^{\rm eff}|=0.2-0.4$ depending on the form
factor  parameterizing $\langle K^-|(\bar s b)_{V-A}|B^-\rangle$;
$|a_2^{\rm eff}|=0.38 \pm 0.05$ is obtained using the form factor
in \cite{Colangelo:1995jv}. This must be compared to $a_2=0.163
(0.126)$ computed  in the naive dimensional regularization (or 't
Hooft-Veltman) scheme \cite{Buras:1998rb}. The difference between
$a_2^{\rm eff}$ and  $a_2$  signals nonfactorizable effects.
However, the clearest evidence of deviation from factorization is
the observation of  $B^- \to K^- \chi_{c0}$, with $\chi_{c0}$ the
lightest $\bar c c$ scalar meson. The data:
\bea{\cal B}(B^- \to K^- \chi_{c0})&=&(6.0^{+2.1}_{-1.8}\pm 1.1)
\times 10^{-4} \;\;\;  \label{belledatum} \\
{\cal B}(B^- \to K^- \chi_{c0})&=&(2.7\pm 0.7) \times 10^{-4}
\,\,\,  \label{babardatum} \eea from BELLE \cite{Abe:2002mw} and
BABAR \cite{Aubert:2002jn}, respectively,
  show that the experimental
amplitude  is non-zero, while the factorized amplitude  vanishes
since $\langle \chi_{c0} |(\bar c c)_{V \mp A}|0\rangle=0$.
Besides, the   rate is comparable to $B^- \to K^- J/\psi$ since,
for example,
 ${{\cal B}(B^- \to K^- \chi_{c0}) \over {\cal B}(B^-
\to K^- J/\psi)}=(0.60^{+0.21}_{-0.18}\pm 0.05\pm0.08)$
\cite{Abe:2002mw}.
\par
 QCD-improved factorization does not reproduce the
measured branching ratios for $B^- \to K^- \chi_{c0}, K^- J/\psi$,
giving either small results or producing infrared divergences, a
signal of uncontrolled nonperturbative effects
\cite{Cheng:2000kt}.
\par
In \cite{Colangelo:2002mj} we investigated  if the deviation from
 factorization  in $B \to c {\bar c}$
 decays may be ascribed to rescattering of
intermediate charm mesons  as in the diagrams  in
fig.\ref{diagrams}. We found that such effects could be large
enough to produce the observed  ${\cal B}(B^- \to K^- \chi_{c0})$.
Other modes with vanishing factorized amplitude
 can  test the rescattering picture,
 as  $B^- \to K^- h_c$ with $h_c$ the lowest lying
$J^{PC}=1^{+-}$ $\bar c c$ state.  $h_c$ is listed by the PDG
among the particles to be confirmed,  with
  $m_{h_c} \simeq 3526 \,MeV$ \cite{Hagiwara:fs}. If $B^- \to K^- h_c$ has
   a sizeable rate, this decay may be used to study 
 $h_c$.
\par
To improve  the  analysis in \cite{Colangelo:2002mj} we introduce
an effective lagrangian for the interactions of all the low-lying
$\ell=1$ $c{\bar c}$ states to   $D^{(*)}_{(s)}$ mesons, based on
the spin symmetry for the heavy quark in the infinite
 mass limit. This  relates all the couplings
to   a single  parameter. The same holds for the couplings of
$\ell=0$ $\bar c c$ mesons to  $D^{(*)}_{(s)}$.

\section{Calculation of  rescattering diagrams}
\label{s:resc} The factorized amplitude $A_F(B^- \to K^- h_c)$ in
(\ref{matrixel}) vanishes since  $\langle h_c |(\bar c c)_{V \mp
A}|0\rangle=0$  due to conservation of parity and charge
conjugation.  However,  the decay can proceed by rescattering
 induced by the same $(\bar b c)(\bar c s)$ effective
Hamiltonian. We consider the process $B^- \to X^0_{\bar u c}
Y^-_{\bar c s} \to K^- h_{ c}$. The lowest lying intermediate
states $X^0_{\bar u c}$,  $Y^-_{\bar c s}$ are $D_s^{(*)-}$,
$D^{(*)0}$ rescattering by  exchange of $D^{(*)}_{(s)}$.
\begin{figure}
\vspace{-0.5cm}
  \resizebox{0.5\textwidth}{!}{\includegraphics{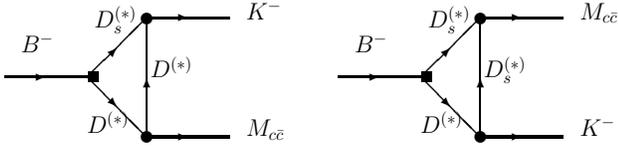}}
  \caption{Typical rescattering diagrams contributing to
$B^- \to K^- M_{c{\bar c}}$, with $M_{c{\bar c}}$ belonging to the
charmonium system.}
\label{diagrams}       
\end{figure}
To compute the diagrams in fig.\ref{diagrams} we need the weak
vertices $B \to D_s^{(*)} D^{(*)}$ and two strong vertices:
 the coupling of  charmed mesons to kaon, and
 of the $h_c$ to  $D_{(s)}^{(*)}$ mesons. All
those vertices are related to few parameters when $m_Q \to
\infty$.
\par
 Interactions of mesons $H_Q$ with a single  heavy
quark $Q$  can be described using the Heavy Quark Effective
Theory, exploiting the heavy quark spin and flavour symmetries
holding for $m_Q \to \infty$. In this limit the heavy quark four
velocity $v$ coincides with that of the hadron and is conserved by
strong interactions  \cite{hqet}. Because of the invariance under
rotations of the heavy quark spin $s_Q$, states differing only for
$s_Q^3$ are degenerate  and form a doublet. When the orbital
angular momentum of the light degrees of freedom relative to $Q$
is $\ell=0$, the states in the doublet have $J^P=(0^-,1^-)$,
corresponding to $(D_{(s)}, D^*_{(s)})$, $(B_{(s)}, B^*_{(s)})$.
The doublet is represented by the matrix $H_a=P_+ [M^\mu_a
\gamma_\mu -M_a \gamma_5]$ with $P_\pm=( 1 \pm \spur{v}) / 2$.
$M^\mu$ is  the vector state, $M$  the pseudoscalar one ($a$ is a
light flavour index). $M_a$, $M^*_a$ contain a factor $\sqrt{m}$,
with $m$ the meson mass.
\par
Let us consider   $B^- \to D_s^{(*)-} D^{(*)0}$, for which
factorization empirically works \cite{Luo:2001mc}. The factorized
amplitude is: \bea \langle D_s^{(*)-} D^{(*)0} | H_W | B^- \rangle
&=& \displaystyle{G_F \over \sqrt{2}}V_{cb}V_{cs}^* a_1 \nonumber
\\ \langle D^{(*)0} | (V-A)^\mu | B^- \rangle &\times&\langle
D_s^{(*)-}| (V-A)_\mu | 0 \rangle \label{fact} \eea with
$a_1=c_1+c_2/N_c$. In the heavy quark limit, the  matrix elements
in (\ref{fact}) can be written in terms of  the Isgur-Wise
function $\xi$, and  a single leptonic constant $\hat F$
\cite{hqet}.
\par
  The $D_s^{(*)} D^{(*)} K$
couplings, in the soft kaon limit, are related to a single
parameter $g$ through the effective lagrangian describing
 strong interactions of charmed mesons with the octet of
the light pseudoscalar mesons \cite{hqet_chir}: $ {\cal L}_I = i
\; g \; Tr[H_b \gamma_\mu \gamma_5 {A}^\mu_{ba} {\bar H}_a]  $. In
${\cal L}_I$: $ {A}_{\mu
ba}=\frac{1}{2}\left(\xi^\dagger\partial_\mu \xi-\xi
\partial_\mu \xi^\dagger\right)_{ba}$,
 ${\bar H}_a=\gamma^0 H_a^\dagger \gamma^0$ and
 $\displaystyle \xi=e^{i {\cal M} \over f}$, with $f\simeq f_{\pi}=131$ MeV and
${\cal M}$ a 3 $\times$ 3 matrix with the light pseudoscalar meson
fields.
\par
We now consider the strong vertex  involving the  $h_c$. For $Q_1
{\bar Q}_2$ mesons   heavy quark flavour symmetry does not hold
any more, but degeneracy is expected under rotations of the two
heavy quark spins. This allows us to build up multiplets for each
value of  $\ell$. For $\ell=0$ one has a $J^P=(0^-,1^-)$ doublet.
The corresponding matrix is \cite{Jenkins:1992nb}: $R^{(Q_1 \bar
Q_2)}=P_+[L^\mu \gamma_\mu -L \gamma_5] P_- $ with $L^\mu=J/\psi$,
$L=\eta_c$ for $\bar c c$. For $\ell=1$, the multiplet is $
P^{(Q_1 \bar Q_2)\mu}= P_+ \Big( \chi_2^{\mu \alpha}\gamma_\alpha
+{1 \over \sqrt{2}}\epsilon^{\mu \alpha \beta \gamma} v_\alpha
\gamma_\beta \chi_{1 \gamma} + {1 \over
\sqrt{3}}(\gamma^\mu-v^\mu) \chi_0 +h_1^\mu \gamma_5 \Big)P_- $.
In the case of $\bar c c$, $\chi_2=\chi_{c2}$, $\chi_1=\chi_{c1}$,
$\chi_0=\chi_{c0}$ correspond to the spin triplet,  $h_1=h_c$ to
the spin singlet \cite{Casalbuoni:1992fd}.
\par
 For $\ell=1$ $Q_1 {\bar
Q}_2$ states, the most general lagrangian describing the coupling
to two heavy-light mesons $Q_1 {\bar q_a}$ and $q_a{\bar Q}_2$ can
be written as follows \cite{Colangelo:2003sa}:
 \bea {\cal
L}_1= i {g_1 \over 2} Tr \left[P^{(Q_1 \bar Q_2)\mu} {\bar H}_{2a}
\gamma_\mu {\bar H}_{1a} \right] + h.c. + (Q_1 \leftrightarrow
Q_2) \,, \nonumber  \eea where   $ H_{2a}= [M^{\prime \mu}_a
\gamma_\mu - M^\prime_a \gamma_5] P_- $ describes $q_a{\bar Q}_2$
mesons. ${\cal L}_1$ is invariant under independent rotations of
the spin of the heavy quarks.
 $g_1$ describes  the interaction of
heavy-light mesons  with the three $\chi_{c}$ states and with
$h_c$, relating the couplings   in absolute value and in sign as
well. This  allows a proper analysis of the diagrams in fig.1.
\par
The interactions of  $\ell=0$ states $R^{(Q_1 \bar Q_2)}$ with the
heavy-light $J^P=(0^-,1^-)$ mesons proceed in P-wave and can be
described by a lagrangian with a derivative: \bea{\cal L}_2=
{g_2\over 2} Tr \left[R^{(Q_1 \bar Q_2)} {\bar H}_{2a}
\stackrel{\leftrightarrow}{{\spur
\partial}} {\bar H}_{1a} \right] +h.c. + (Q_1 \leftrightarrow Q_2)
\nonumber \eea again invariant under heavy quark spin rotations.
\par
  $g_1$ and $g_2$ can be
estimated using vector meson dominance (VMD). One can consider
$\displaystyle \langle D(v^\prime) | \bar c c | D(v) \rangle$,
assuming the dominance in the $t$-channel of the
 $0^+$ $\bar c c$ state, and using the
normalization  $\xi(1)=1$. One obtains $ g_{D D \chi_{c0}}= 2
{\displaystyle m_D m_{\chi_{c0}} / f_{\chi_{c0}}}$, where
$f_{\chi_{c0}}$ is defined by $\langle 0| \bar c c |
\chi_{c0}(q)\rangle= f_{\chi_{c0}} m_{\chi_{c0}}$. This relation
gives: $ g_1=-\sqrt{ m_{\chi_{c0}}}/( \sqrt{3} f_{\chi_{c0}})$.
The same argument gives $g_2=\displaystyle{\sqrt {m_\psi} /( 2 m_D
f_\psi)} $, where
 $f_\psi$ is defined by $\langle 0| \bar c
\gamma^\mu c | J/\psi(p,\epsilon)\rangle= f_\psi m_\psi
\epsilon^\mu$.  To  account for the off-shell effect of the
exchanged particles, the virtuality of which can be large, we
write: $g_i(t)=g_{i0}\,F_i(t)$, with $g_{i0}$ the on-shell
couplings and: $ F_i(t)=(\Lambda_i^2 -m^2_{D^{(*)}}
)/(\Lambda_i^2-t)$. The parameters $\Lambda_i$ represent a source
of uncertainty.
\par
In the numerical analysis
 we exploit the heavy quark limit,
putting $f_{D^*_s}=f_{D_s}$ and use $f_{D_s}=240$ MeV
\cite{Colangelo:2000dp}. For the Isgur-Wise function, we use
$\xi(y)= \left( {2 /( 1+y)} \right)^2$, compatible with data from
 semileptonic $B \to D^{(*)}$ decays. As for  $g$,  CLEO
Collaboration obtained: $g=0.59 \pm0.01 \pm 0.07$ by  measuring
$\Gamma(D^*)$  and the $D^*$ branching fraction to $D \pi$
\cite{Anastassov:2001cw}. This value should be compared to
theoretical estimates ranging from $g\simeq 0.3$ up to $g \simeq
0.77$ \cite{gnew}. Since  the rescattering amplitudes always
depend on $g \cdot F_i(t)$, we put $\Lambda_i=\Lambda$ for all $ i
$ and use $g=0.59$, leaving to $\Lambda$ the task of spanning the
range of possible variation of $g$.
\par
For $g_1$ and $g_2$ we use the VMD relations, together with the
QCD sum rule result $f_{\chi_{c0}}=510 \pm 40 $ MeV
\cite{Colangelo:2002mj} and  the experimental value
$f_{J/\psi}=405 \pm 14 $ MeV. \par  We computed the imaginary part
of the rescattering diagrams; the determination of the real part
is more uncertain. Using a dispersive representation,
 we obtained for  $B^- \to K^-
\chi_{c0}(J/\psi)$ that Re${\cal A}_i \simeq $Im${\cal A}_i$
\cite{Colangelo:2002mj}. Hence we account for the real part of the
amplitudes considering them as fractions of the imaginary part
varying from $0$ to $100 \%$. Such an uncertainty  will affect the
final result.
\par
We constrain   $\Lambda$ considering rescattering contributions to
$B^- \to K^- J/\psi$, where  ${\cal A}(B^- \to K^- J/\psi)={\cal
A}_{fact}+{\cal A}_{resc}$ is bounded by  experimental data. For
$\Lambda \simeq  2.6-3$ GeV, the sum does not exceed the
experimental bound. Moreover, we repeat the analysis in
\cite{Colangelo:2002mj} for $B^- \to K^- \chi_{c0}$,  using now
the relations stemming from the  lagrangian ${\cal L}_1$. We get a
branching ratio compatible with the experimental result from BABAR
if $\Lambda$ is varied around $3.0$ GeV.
\par
\begin{figure}\vspace{-1.5cm}
  \resizebox{0.43\textwidth}{!}{\includegraphics{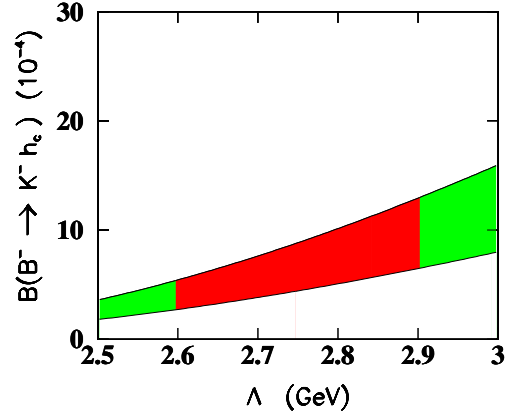}}\vspace{-1cm}
  \caption{Branching ratio ${\cal B}(B^- \to K^- h_{c})$ versus
 $\Lambda$. The lowest curve corresponds to ${\rm Re}
{\cal A}_i=0$, the highest one to ${\rm Re} {\cal A}_i={\rm Im}
{\cal A}_i$. The dark region corresponds to the result
(\ref{result-hc}).}
\label{br_hc}       
\end{figure}

Using such constraints  we analyze  $B^- \to K^- h_c$. In
fig.\ref{br_hc} we plot the branching ratio   versus $\Lambda$. We
find \cite{Colangelo:2003sa}: \be {\cal B}(B^- \to K^- h_c)=(2 -
12) \times 10^{-4} \,\,\, , \label{result-hc} \ee where we
accounted for the uncertainty on the real part of the amplitudes
and on the variation of  $\Lambda$. This result suggests  ${ \cal
B} (B^- \to K^- h_c)$  large enough to be measured at the
B-factories. Moreover, this mode  represents a sizeable fraction
of the inclusive $B^- \to X h_c$ decay, for which, considering the
production of the $c{\bar c}$ pair in $h_c$ in the color-octet
state,  ${\cal B}(B^- \to  h_c X)=(13 - 34) \times 10^{-4}$ is
predicted \cite{Beneke:1998ks}.

The main uncertainty affecting our results
 is due to  cancellations between different
amplitudes, which are of similar size.  Another uncertainty is due
to the neglect of contributions of higher  states,  even though a
minor role can be presumed for higher resonances.
\par
Bearing such uncertainties in mind  we can conclude that
rescattering terms may contribute to the nonfactorizable effects
observed in $B \to $ charmonium transitions.

\section{Conclusions}\label{s:conc}
The $h_c$ was  observed in $p{\bar p}$ annihilation
 and in $p-{\rm Li}$
interactions \cite{Hagiwara:fs}. In $B$ decays, one could access
$h_c$ looking either at its hadronic modes: $h_c \to J/\psi
\pi^0$, $\rho^0 \pi^0$, etc. ,  or at its radiative modes: $h_c
\to \eta_c \gamma$, $\chi_{c0} \gamma$, etc. The channel $h_c \to
\eta_c \gamma$ seems promising, as noticed by Suzuki who
estimated: ${\cal B}(h_c \to \eta_c \gamma) \simeq 0.50 \pm 0.11$
\cite{Suzuki:2002sq}. A similar result:  ${\cal B}(h_c \to \eta_c
\gamma) = 0.377$
 is obtained in \cite{Godfrey:2002rp}. These
two predictions, together with the experimental  ${\cal B}(\eta_c
\to K {\bar K} \pi)$ and our result (\ref{result-hc}), allow us to
predict: ${\cal B}(B^- \to K^- h_c \to K^- \eta_c \gamma \to K^- K
{\bar K} \pi \gamma)= (4 - 26) 10^{-6} $,  within the reach of
current experiments.
\par
As for  rescattering effects in $B \to $ charmonium decays, we
 found that they can  produce  a branching ratio for $B^- \to
K^- h_c$ comparable with that of $B^- \to K^- \chi_{c0}$. The same
holds  for  $B^- \to K^- \psi(3770)$ which, because of the
smallness of $f_{\psi(3770)}$, is predicted by factorization with
a tiny branching ratio. The experimental measure ${\cal B}(B^- \to
K^- \psi(3770))=(0.48\pm0.11\pm0.12)\times 10^{-3}$
\cite{Abe:2003zv} is a further evidence of  large nonfactorizable
effects. In our approach, using $g_{D D \psi(3770)}=14.94 \pm
0.86$ obtained from the width of $\psi(3770)$, we  get ${\cal
B}(B^- \to K^- \psi(3770))=(0.9-4)\times 10^{-4}$. Similar
conclusion applies to $B^- \to K^- \chi_{c2}$ with $\chi_{c2}$ the
$J^{PC}=2^{++}$ $c{\bar c}$ state, the factorized amplitude of
which also vanishes. The observation of this process with
branching ratio comparable to ${\cal B}(B^- \to K^- \chi_{c0})$,
 ${\cal B}(B^- \to K^- h_c)$  would support the rescattering
picture.
\par
{\bf Acknowlwdgments} I thank P. Colangelo, T.N. Pham for
collaboration. Partial support from the EC Contract No.
HPRN-CT-2002-00311 (EURIDICE) is acknowledged.
%

\end{document}